\documentclass[aps,prx,reprint,preprintnumbers,superscriptaddress,nofootinbib,longbibliography,floatfix]{revtex4-2}
\pdfoutput=1
\usepackage{rotating}
\usepackage{array}
\usepackage{amsmath}
\usepackage[normalem]{ulem}
\usepackage{slashed}
\usepackage{booktabs}
\usepackage[pdftex,table]{xcolor}
\usepackage{units}
\usepackage{xfrac}
\usepackage{mathtools}
\usepackage{empheq}
\usepackage[]{units}
\usepackage{multirow}
\usepackage{amssymb}
\usepackage{url}
\usepackage{comment}
\usepackage{physics}
\usepackage{color,soul}
\usepackage{bbm}
\usepackage[caption=false]{subfig}
\usepackage{adjustbox}
\usepackage[T1]{fontenc}

\usepackage{hyperref}
\hypersetup{
  colorlinks=true,
  citecolor=blue,
  linkcolor=blue,
  urlcolor=blue
}

\newcommand{\pt}{\mathrm{p_{T}}}

\newcommand{\omni}{\textsc{OmniLearn}~}

\begin{document}

\title{Foundation Model Framework for All Tasks Involving Jet Physics}
\author{Wahid Bhimji}
\affiliation{NERSC, Lawrence Berkeley National Laboratory, Berkeley, CA 94720, USA}

\author{Chris Harris}
\affiliation{NERSC, Lawrence Berkeley National Laboratory, Berkeley, CA 94720, USA}

\author{Vinicius Mikuni}
\email{vmikuni@hepl.phys.nagoya-u.ac.jp}
\affiliation{Nagoya University, Kobayashi-Maskawa Institute, Aichi 464-8602, Japan}

\author{Benjamin Nachman}
\email{nachman@stanford.edu}
\affiliation{Department of Particle Physics and Astrophysics, Stanford University, Stanford, CA 94305, USA}
\affiliation{Fundamental Physics Directorate, SLAC National Accelerator Laboratory, Menlo Park, CA 94025, USA}

\begin{abstract}
Foundation models use large datasets to build an effective representation of data that can be deployed on diverse downstream tasks.  Previous research developed the \textsc{OmniLearn} foundation model for jet physics, using unique properties of particle physics, and showed that it could significantly advance discovery potential across collider experiments.  This paper introduces a major upgrade, resulting in the \textsc{OmniLearned} framework.  This framework has three new elements: (1) updates to the model architecture and training, (2) using over one billion jets used for training, and (3) providing well-documented software for accessing all datasets and models.  We demonstrate \textsc{OmniLearned} with three representative tasks: top-quark jet tagging with the community Delphes-based benchmark dataset, b-tagging with ATLAS full simulation, and anomaly detection with CMS experimental data.  In each case, \textsc{OmniLearned} is the state of the art, further expanding the discovery potential of past, current, and future collider experiments. 
%
\end{abstract}

\maketitle

\vspace{10mm}

\section{Introduction}
\label{sec:intro}

Hadronic jets are the result of high-energy quarks and gluons.  These objects are ubiquitous in reactions at colliders and elsewhere that involve momentum transfers well above 1 GeV.  The radiation pattern within jets encodes information about their origin as well as emergent properties of the strong force.  For decades, researchers proposed physically-inspired observables to characterize this complex structure~\cite{Kogler:2018hem,Larkoski:2017jix,Marzani:2019hun}.  Deep learning-based taggers are now standard, but it is challenging to assemble enough examples jets to train state-of-the-art machine learning models for each of the myriad tasks involving jets. 

A solution to this challenge is to pretrain an expressive machine learning model on a large number of related examples and then fine tune the model for downstream tasks.  Since the learned representation can be used for many goals, it serves as a foundation model.  Most research related to foundation models in particle physics have focused on tokenized models trained via self-supervised learning~\cite{Feickert:2021ajf,Birk:2024knn,Hallin:2025ywf,Harris:2024sra,Golling:2024abg,Leigh:2024ked,Bardhan:2025icr}, the same setup used to build large language models (LLMs).  These approaches learn useful representations, but do not make full use of the continuity of the data and/or the label information available from simulations.

An alternative approach is to train using classification and generation objectives that are similar in nature to the ones employed by typical downstream tasks.  This is the core philosophy of the \textsc{OmniLearn} foundation model for jets~\cite{Mikuni:2025tar,Mikuni:2024qsr}.  For that model, a transformer-based architecture was trained to classify and generate jets from the 100-million jet \textsc{Delphes}~\cite{deFavereau:2013fsa} fast-simulation-based JetClass dataset with 10 classes~\cite{Qu:2022mxj}.  The model was then fine tuned on a number of downstream tasks using Geant~\cite{GEANT4:2002zbu}-based full  simulations from ATLAS, CMS, and H1.  In all cases, \textsc{OmniLearn} matched or exceeded the state of the art.  For example, \textsc{OmniLearn} can significantly reduce the number of computationally-expensive, fully-simulated examples required to train jet taggers, can accelerate and stabilize neural likelihood-ratio estimation, and can facilitate full phase-space anomaly detection.

In this paper, we present a significant upgrade of \textsc{OmniLearn}.  This includes a number of updates to the architecture and training to improve performance.  These updates also allow for more per-jet-constituent information to be incorporated into the model as well as for datasets without labels to be used in the foundation model training.  The upgraded model, called \textsc{OmniLearned}, is now trained with over one billion jets (10x the previous model).  This is a significant milestone on its own and utilizes the scaling properties of transformers.  Alongside this paper, we have prepared software for readily accessing all of the train and test datasets in a unified framework as well as well-documented code for training/fine-tuning the machine learning models.  The new framework is demonstrated on three key tasks.  First, the models are compared with the current state of the art on the community benchmark top quark and quark-versus-gluon tagging datasets.  Second, we are able to improve the performance of $b$-tagging with ATLAS public simulations.  As part of this task, we also show how to repurpose parts of the foundation model for tasks different than their original intention (per-track classification from jet generator).  Lastly, we show how the top quark can be effectively re-discovered using anomaly detection deployed on CMS open data.  Along the way, we show how the outputs of the foundation model can also be used directly (without fine tuning) for anomaly detection when the target is relatively close to a class in the training dataset.


The paper is organized as follows. Sec.~\ref{sec:pet} describes the improvements to the Point Edge Transformer (\textsc{PET}) model that forms the backbone of \textsc{OmniLearned}. Sec.~\ref{sec:dataset} then introduces the dataset used to train \textsc{OmniLearned} with over 1 billion jets. Results for different tasks are then described in Sec.~\ref{sec:top_qg} for classification using several benchmarks, and Sec.~\ref{sec:ad} on performing anomaly detection using the CMS Open Data.  The paper ends with conclusions and outlook in Sec.~\ref{sec:conclusions}.

\section{OmniLearned}
\label{sec:pet}

\begin{figure*}[ht]
    \centering
        \includegraphics[width=.95\textwidth]{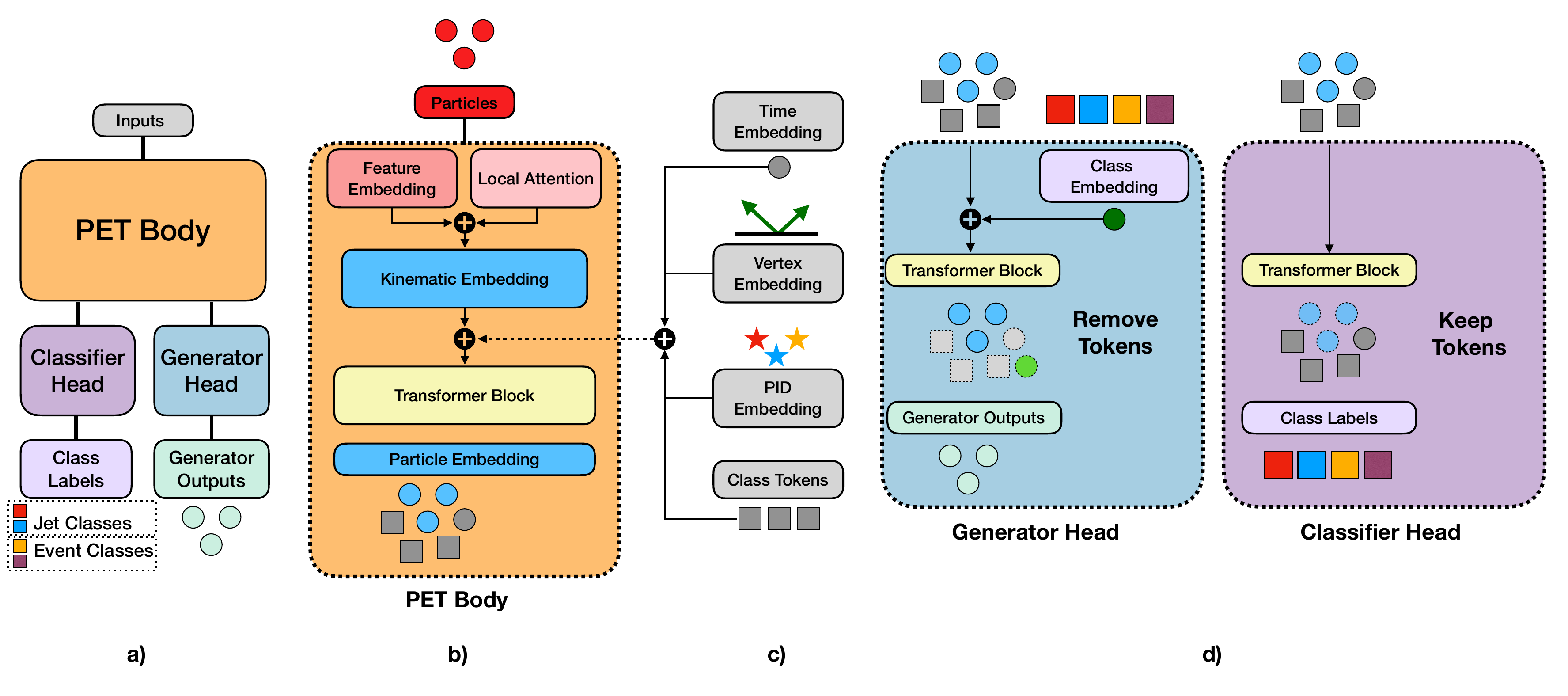}
    \caption{Neural network architecture used to train \textsc{OmniLearned}. The general model architecture (a) consists of the new \textsc{PET v2} body (b), input specific embedding (c) and task-specific blocks (d). See the text for more details.}
    \label{fig:pet}
\end{figure*}

In the previous \omni model, a transformer network combined with a graph neural network named point-edge transformer (\textsc{PET}) was used as the backbone architecture. This choice naturally addresses the data structure of jets as point clouds with varying number of particles, an approach that was successfully used for multiple tasks in collider physics~\cite{Komiske:2018cqr,ATL-PHYS-PUB-2020-014,Moreno:2019bmu,Qu:2019gqs,Shimmin:2021pkm,Mikuni:2020wpr,Mikuni:2021pou,Qu:2022mxj}. Building on the success of the \textsc{PET} model, we further improve the architecture (to \textsc{PET v2}) by modifying the following key properties to enhance expressiveness and flexibility to the model. The new model architecture is shown in Fig~\ref{fig:pet}.

\subsection{Input Features}
In the previous \omni implementation, the inputs to the model  were the kinematic information of each particle clustered in the jet. While only four numbers are necessary to completely describe the momentum of each particle, redundant features were added to accelerate training convergence.  Such features included the difference in pseudorapidity $\eta$ and azimuthal angle $\phi$ between each particle and the jet axis, together with the absolute distance $\Delta_R$, calculated using the same $\eta$ and $\phi$ values. In \textsc{OmiLearned}, we maintain the difference in $\eta$ and $\phi$ between the particle and jet axis as inputs but only add two additional features, consisting of the logarithm of the transverse momentum $\log p_{T}$ and energy $\log E$. We also remove the pre-processing step, previously used to standardize the mean and variance of the input features, since this choice of representation already leads to each input roughly centered at zero with $\mathcal{O}(1)$ variance. That leaves us with the minimal set of inputs supported by \textsc{OmniLearned} as $(\Delta\eta,\Delta\phi,\log p_{T}, \log E)$.

\subsection{Encoding Additional Information}

While the kinematic information for each particle is always present in all datasets considered in this work, some datasets also provide additional information such as particle identification (PID) or vertex information for charged particles. In the previous \omni model, we addressed this challenge by creating specific embedding blocks to address additional features. During training, the additional features would be randomly dropped with fixed probability and all additional features were replaced by zeros. This strategy is similar to dropout~\cite{JMLR:v15:srivastava14a} and allows the model to learn a useful representation in the presence and absence of these additional features. We maintain the same successful methodology but modify the exact implementation of each network block. First, the PID information is handled separately from the vertex information. Since the PID is naturally a discrete feature used to label a particle as an electron, muon, photon, charged or neutral hadron, we use a simple lookup table to embed the PID information, similarly to how tokenized words are encoded in language models. The vertex information is also handled separately from the other features and is encoded using two MLP layers with \textsc{GELU}~\cite{Goriely:2022upe} non-linearity. The result of the encoded PID and vertex features are added to the kinematic features such that in the absence of either PID or vertex information the corresponding encoding is set to zero. When adapting the model to new datasets, if the dataset contains either PID or vertex information, the corresponding encoding blocks are loaded from the base model; otherwise they are safely ignored. The time parameter used to condition the diffusion generation (more in Sec.~\ref{sec:loss}) is also embed in the model using Fourier features~\cite{tancik2020fourier} followed by MLP layers with \textsc{GELU} activation. Instead of adding the time embedding to the overall latent representation we include the time information as an additional point in the point cloud. This choice, paired with the use of transformer blocks, improves the conditioning of the diffusion model with the time parameter, improving the generation quality during downstream tasks. 

\subsection{Local Attention Layers}
The use of local information improves the ability of the model to grasp correlations between particles inside the jets. These local correlations can be incorporated using implementations similar to dynamic graph convolution (DGCNNs) layers where the neighborhood is defined using a $k$-nearest neighbor algorithm. For each particle $x_i$ in the jet, we calculate pairwise features $f(x_i,x_j)$ with respect to each of the j-neighbors. In the original \omni model, the function $f(x_i,x_j) = [x_i, x_i - x_j] $ was chosen. Multilayer perceptron (MLP) layers then used the pairwise features as inputs before the average across all neighbors was calculated. While this implementation has been successful in encoding the local relationship between particles, we perform two modifications that further improve the expressiveness and scalability of these operations. First,  inspired by Ref.~\cite{Wu:2024thh} we include physics-inspired interaction terms between each particle and their $k$-nearest neighbors. In particular, we modify the function $f(x_i,x_j)$ to be:

\begin{align}
    f(x_i,x_j) = [&x_i - x_j, \log m(x_i,x_j), \log\Delta_R(x_i,x_j), \notag\\
                 &\log[\min \{p_{Ti}, p_{Tj}\}\times\Delta_R(x_i,x_j)]],
    \label{eq:local_features}
\end{align}
where $m(x_1,x_j)$ is the invariant mass of the four-vector sum  and $\Delta_R(x_i,x_j)$ is the distance in the pseudo-rapidity-azimuthal plane between particles $x_i$ and $x_j$ . The second modification is to replace the simple average with a learnable weighted average. In practice, we implement a transformer block where the attention mechanisms attends to only particle $x_i$ and their $k$-neighbors. Since attention is a computationally expensive operation, we set $k = 10$, which leads to the overall complexity of $N\times k \times k$. Since we later consider jets with up to 150 particles, the final complexity of the local attention layers is roughly equivalent to the one obtained by having a transformer block where all particles attend to all other particles. The use of attention layers instead of a simple average allows the model to learn the importance of each neighbor when creating the local embedding while also becoming more robust against variations in the overall particle multiplicity in the jet.

\subsection{Global Attention Layers}

The use of transformer layers~\cite{DBLP:journals/corr/VaswaniSPUJGKP17} is now the default architecture used both in collider physics and elsewhere due to the scalability of transformer models when trained with large amounts of data. In the previous \textsc{PET} implementation, a generic transformer block was used.  In the new implementation, we modify the attention mechanism to add the same set of physics inspired interaction terms to the attention matrix. Effectively, we modify the attention matrix by adding a bias term that encodes the same physics quantities from Eq.~\ref{eq:local_features}, now calculated between each pair of particles. In addition, we replace the standard layer normalization operations with a learnable hyperbolic tangent operation~\cite{zhu2025transformers} which is shown to improve stability even when considering dozens of transformer blocks. 

\subsection{Task Specific Heads}

The joint data representation is used as input to task specific blocks that aim to solve complementary tasks in the form of jet classification and jet generation. In the previous \omni implementation, the classification head used a classification token, implemented as a learnable token tasked to summarize the information of all particles inside the jet. This additional token was treated in the same way as the other particles in the jet and attended to all particles during transformer blocks. We noticed that adding more learnable tokens was helpful, resulting in each token learning to summarize different particle information, similar to a multi-head algorithm. In our new implementation, we use five learnable tokens that are now added to the set of particles as part of the general embedding model instead of only being added at the classification block. For jet classification, the tokens themselves are used to determine the classification outputs, while for jet generation, the additional tokens are removed before the output layer.

\subsection{Loss Function}
\label{sec:loss}

The loss introduced in \omni consisted of a combination of the classification and generation tasks. To combine the two representations, an additional loss term took the perturbed inputs, designed for the training of the diffusion models, through the classification head. The overall loss function was:

\begin{align}
    \mathcal{L} &= \mathcal{L}_{\text{class}} + \mathcal{L}_{\text{gen}} + \mathcal{L}_{\text{class smear}} \\
                &= \textsc{CE}(y,y_{\text{pred}}) +  \left\| v - v_{\text{pred}}\right\|^2 + \alpha^2\textsc{CE}(y,\hat{y}_{\text{pred}})\,.
    \label{eq:loss}
\end{align}
The cross-entropy (CE) term is calculated using the true jet labels and the output of the classification head. In the new implementation, we would also like to include jets where the truth label is not present, which is the case for data collected by experiments. In the absence of truth labels, we design a separate classification task, where the goal is to classify jets coming from different datasets. In this case, we break down the output of the classifier head into two, one used for supervised jet flavor classification, and one used for sample classification, where the cross entropy loss for both classification tasks are added equally in the loss function.

We also modify the training of the generative part based on diffusion models. During the diffusion model training, particles clustered inside the jet were first perturbed using a time-dependent Gaussian value:

\begin{equation}
    z(t) = \alpha(t)x + \sigma(t)\epsilon,
    \label{eq:diffusion}
\end{equation}
where $\epsilon\sim\mathcal{N}(0,1)$ and $\alpha$ and $\sigma$ determine the mixture between clean data $x$ and noise $\epsilon$. The goal of the diffusion model is then to take as input $z(t)$ and predict the velocity term $v = \alpha(t)\epsilon -\sigma(t)x$ introduced in Ref.~\cite{salimans2022progressive}, a linear combination of the noise and the clean data. This approach has been very successful at generating jets~\cite{PhysRevD.108.036025,Araz:2024bom} with high fidelity. Recently, an alternative parameterization has been introduced in the form of flow matching models~\cite{lipman2022flow}, where the perturbed data is used as inputs to a model that instead predicts the so-called \textit{velocity field} $u = \epsilon - x$. The velocity field estimates the path that connects the noise distribution to the data as the solution of an ordinary differential equation. Even though diffusion and flow matching models are presented as different philosophies, they are equivalent. In fact, for the choice of $\alpha(t) = 1 - t$ and $\sigma(t) = t$, often used in flow matching models, the perturbed data used as input for the flow matching training is the same as the one from Eq.~\ref{eq:diffusion}. Moreover, the diffusion loss function defined as:

\begin{equation}
    \mathcal{L}_D = \left\| v_\theta(z_t,t) - v \right\|^2,
\end{equation}
for a neural network model $v_\theta(z_t,t)$ with trainable parameters $\theta$ is related to the flow matching loss function

\begin{equation}
    \mathcal{L}_F =  \left\| u_\theta(z_t,t) - u \right\|^2 = w(t)\left\| v_\theta(z_t,t) - v \right\|^2,
    \label{eq:fm}
\end{equation}
 through a time-dependent weight $w(t) = \left [ \frac{\sigma + \alpha}{\sigma^2 + \alpha^2} \right ]^2$.

Even though both formulations lead to similar losses, the training convergence rate and sample quality between diffusion and flow matching models can be different, with flow matching models often leading to better generation quality with similar number of sampling steps as the diffusion training. For this reason, we change the velocity prediction from \omni to the flow matching objective in Eq.~\ref{eq:fm} resulting in the combined loss:

\begin{align}
    \mathcal{L} = \textsc{CE}(y,y_{\text{pred}}) +  \left\| u - u_{\text{pred}}\right\|^2 + \alpha^2\textsc{CE}(y,\hat{y}_{\text{pred}})\,.
    \label{eq:loss_update}
\end{align}

This change also leads to the modification to the function $\alpha$ in Eq.~\ref{eq:loss} that multiplies the cross entropy term evaluated over perturbed inputs.

\section{1 Billion Dataset}
\label{sec:dataset}

The growth of the machine learning community for high energy physics closely follows the increase in availability of open datasets used for benchmarking and research. Multiple datasets for jet physics have been made available, each hosted on different platforms and saved with different data formats. As part of the \textsc{OmniLearned} release, we also provide a software package that automatically accesses and downloads all datasets used in this work to unify them in the same format. The datasets used for the pre-training of \textsc{OmniLearned} are listed in Tab.~\ref{tab:datasets}.

\begin{table}[th]
    \centering
    \caption{List of open datasets used during the pre-training of \textsc{OmniLearned} and the amount of events available for training, testing, and validation.}
    \label{tab:datasets}
	\begin{tabular}{lccccc}
    \hline
          Dataset &  Training & Validation & Test \\
            \hline
            JetClass~\cite{Qu:2022mxj} & 100M  & 20M & 5M \\ 
            JetClass2~\cite{Li:2024htp} & 200M  & 600k & 600k \\ 
            Aspen Open Jets~\cite{Amram:2024fjg} & 125M  & 25.7M & 26.6M \\ 
            ATLAS Top Tagging~\cite{ATLAS:2024rua} & 178M  & 20M & 2.2M \\ 
            H1 DIS & 42.2M  & 872k & 255k \\ 
            CMS QCD & 239M  & 17.5M & 16M \\ 
            CMS BSM & 173.5M  & 17M & 17M \\ 
            \hline
            Total & 1057.7M  & 101.8M & 67.6M \\ 
            
	\end{tabular}
\end{table}

The combined dataset size amounts to more than 1 billion jets. A total of 210 classes are created across all datasets, 200 of them targeting specific jet flavors and 10 of them attached to specific datasets. The first dataset included in the pre-training is the JetClass dataset~\cite{Qu:2022mxj} consisting of 10 different jet classes simulated using MADGRAPH5\_aMC@NLO~\cite{Alwall:2014hca} for the matrix element calculation and \textsc{Pythia 8}~\cite{Sjostrand:2006za,Sjostrand:2014zea} for parton showering and hadronization. \textsc{Delphes 3.4.3}~\cite{deFavereau:2013fsa,Mertens:2015kba,Selvaggi:2014mya} is used to simulate detector effects with the CMS detector configuration. Jets are clustered using the anti-$k_t$ algorithm with radius parameter of $R=0.8$~\cite{Cacciari:2005hq,Cacciari:2011ma,Cacciari:2008gp}. Only jets with transverse momentum between 500-1000 GeV and pseudorapidity $|\eta| < 2.0$ are considered. The  JetClass dataset was upgraded with additional 188 classes in the JetClass 2 dataset~\cite{Li:2024htp}, consisting of fine-grained labels for different combinations of initial partons, but using the same generators, detector simulation, and jet clustering setup as JetClass. 
The ATLAS Top Tagging dataset~\cite{ATLAS:2024rua} released by the ATLAS Collaboration consists of top quarks and Quantum Chromodynamics (QCD) non-resonant jets simulated using the ATLAS detector simulation based on Geant4~\cite{GEANT4:2002zbu,ATLAS:2010arf}.  Events are generated with \textsc{Pythia8} using the NNPDF2.3LO~\cite{Ball:2012cx} set of parton distribution functions and the A14~\cite{Buckley:2014ctn} set of tuned parameters. Pileup effects are simulated by overlaying inelastic interactions on top of the underlying hard scattering process based on the 2017 data taking period. Jet constituents are reconstructed based on the Unified Flow Object algorithm~\cite{ATLAS:2020gwe}. Jets are clustered using anti-$k_t$ algorithm with R=1.0 with additional Soft-Drop~\cite{Larkoski:2014wba} and pileup mitigation ~\cite{Berta:2014eza,Berta:2019hnj,Cacciari:2014gra} algorithms applied. Compared to the previous version used by \textsc{OmniLearn}, the ATLAS Top Tagging dataset was updated to also contain systematic variations of the nominal samples used for the simulations. These variations cover the same set of uncertainties commonly considered for physics analysis at the LHC. Since we only use this dataset for the pre-training of the model, we include both nominal and varied datasets together during training, without explicitly making a distinction between systematic variations and nominal samples. These events also receive the same label assigned for top quarks and QCD events in the JetClass dataset for consistency.
To increase the variety of jet examples at different regions of the phase space we also include jets simulated across different collision systems. We use simulations of neutral-current deep inelastic scattering (DIS) provided by the H1 Collaboration and generated using two different generators: the Rapgap 3.1~\cite{Jung:1993gf} and Djangoh 1.4~\cite{Charchula:1994kf} generators for electron-proton collisions with electron and proton beam energies of 27.6 GeV and 920 GeV, respectively. The detector simulation of the H1 detector is performed using the Geant3~\cite{Brun:1987ma, Britzger:2021xcx} package. An energy-flow algorithm~\cite{energyflowthesis,energyflowthesis2,energyflowthesis3} is then used to reconstruct the particles clustered into jets using the $k_t$ algorithm with R=1.0. Since these events target a different physics process, we assign distinct labels to each simulation, separately from the other labels considered so far. 
The last set of events used for the pre-training of \textsc{OmniLearned} are processed from the open dataset released by the CMS Collaboration, consisting of teal or simulated proton-proton collisions collected in 2016. The preprocessed data events were also released as part of the Aspen Open Jets~\cite{Amram:2024fjg} dataset. Additional simulations from the CMS Collaboration for QCD and several potential new physics processes were also produced using the same framework. The new physics samples were chosen such that they individually yield a large amount of jets in the final state, while providing complementary information based on the datasets already considered.  The selected samples are vector-like quark production with a B$^\prime$ and T$^\prime$, charged Higgs production, bulk Graviton and radion production, supersymmetry (SUSY) in the context of Next to Minimal Supersymmetric Standard Model (NMSSM), SUSY with displaced vertices, Z$^\prime$ production, and a narrow resonance X that decays to a resonant pair YY which in turn decay to pairs of jets. These new physics samples combined with the QCD and the experimental data collected by the CMS Collaboration form the additional 10 classes used for sample classification during the pre-training phase.

In all  datasets, up to 150 particles are saved per jet to be used during training. The training is carried out on the Perlmutter Supercomputer~\cite{Perlmutter} using from 32 to 512 GPUs simultaneously. For different model sizes, the local batch was chosen based on the memory available for each GPU. The global batch size, defined as the local batch size multiplied by the number of GPUs, is kept the same for all models and equal to 4096. All pre-trained models are trained for 3 full passes of the 1 billion dataset. \textsc{OmniLearned} is implemented in \textsc{Pytorch}~\cite{pytorch}. The cosine learning rate schedule~\cite{DBLP:journals/corr/LoshchilovH16a} is used with an initial learning rate of $1\times10^{-5}$ and decreased to $10^{-6}$ until the end of the training. The \textsc{Lion} optimizer~\cite{chen2024symbolic} is used with parameters $\beta_1 = 0.95$ and $\beta_2 = 0.98$. The fine-tuning of \textsc{OmniLearned} across different datasets and tasks is performed by setting the learning rate of all network weights to be a factor 5 smaller than the output layer. We investigate three model sizes for \textsc{OmniLearned} named small (-s), medium (-m), and large (-l) models with roughly 3M, 58M and 460M trainable weights respectively. In App.~\ref{app:models} the details of each model size is provided.  Some hyperparameter optimization of all parameters stated above suggested that large gains are not readily available from modest tweaks to the setup.

\section{Jet Classification}
\label{sec:top_qg}

\begin{figure*}[ht]
    \centering
        \includegraphics[width=.95\textwidth]{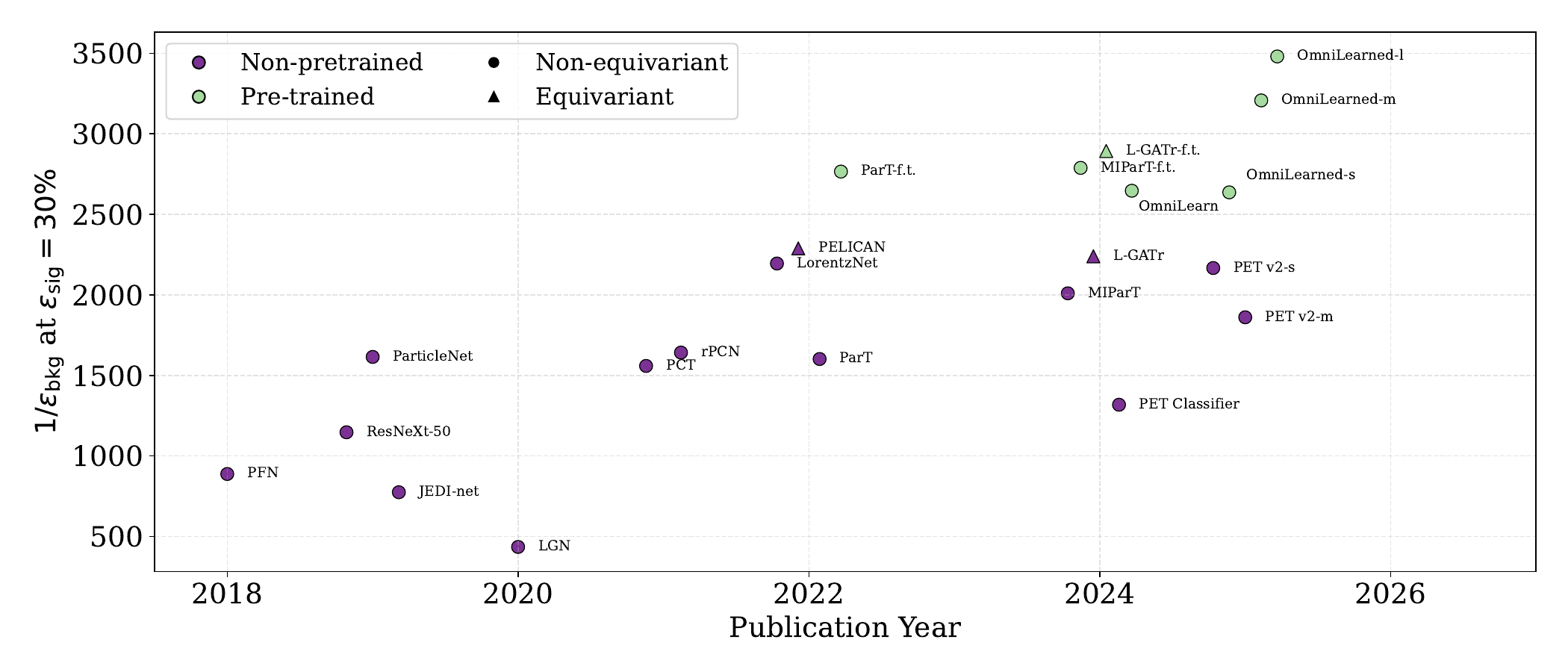}
    \caption{Background rejection efficiency for a fixed signal efficiency of $30\%$ in the community top tagging dataset.}
    \label{fig:top_tagging}
\end{figure*}

\begin{table}[th]
    \centering
    \caption{Comparison between the performance reported for different classification algorithms on the top tagging dataset. The uncertainty quoted corresponds to the standard deviation of five trainings with different random weight initialization. If the uncertainty is not quoted then the variation is negligible compared to the expected value. Bold results represent the algorithm with highest performance. }
    \label{tab:results_top}
	\begin{tabular}{lccccc}
    \noalign{\smallskip}\hline
          &  Acc &AUC & \multicolumn{2}{c}{1/$\epsilon_B$} \\
          \cline{4-5}
          & & & $\epsilon_S = 0.5$ & $\epsilon_S = 0.3$ \\
            \hline
            ResNeXt-50~\cite{Qu:2019gqs} & 0.936 & 0.9837 & 302 $\pm$ 5 & 1147 $\pm$ 58 \\
            P-CNN~\cite{Qu:2019gqs} & 0.930 & 0.9803 & 201 $\pm$ 4 & 759 $\pm$ 24 \\
            PFN~\cite{Komiske:2018cqr} & - & 0.9819 & 247 $\pm$ 3 & 888 $\pm$ 17 \\
            ParticleNet~\cite{Qu:2019gqs} & 0.940 & 0.9858 & 397 $\pm$ 7 & 1615 $\pm$ 93\\
            JEDI-net~\cite{Moreno:2019bmu}  & 0.9300 & 0.9807 & - & 774.6 \\
            PCT~\cite{Mikuni:2021pou} & 0.940 & 0.9855 & 392 $\pm$ 11 & 1559 $\pm$ 98 \\
            LGN~\cite{Bogatskiy:2020tje} & 0.929 & 0.964 & - & 435 $\pm$ 95\\
            rPCN~\cite{Shimmin:2021pkm} & - & 0.9845 & 364 $\pm$ 9 & 1642 $\pm$ 93\\
            LorentzNet~\cite{Gong:2022lye} & 0.942 & 0.9868 & 498 $\pm$ 18 & 2195 $\pm$ 173 \\
            PELICAN~\cite{Bogatskiy:2022czk} & 0.9425 & 0.9869 & - & 2289 $\pm$ 204  \\
            ParT~\cite{Qu:2022mxj} & 0.940 & 0.9858 & 413 $\pm$ 16 &  1602 $\pm$ 81 \\
            ParT-f.t.~\cite{Qu:2022mxj} & \textbf{0.944 }&  0.9877 & \textbf{691 $\pm$ 15} & 2766 $\pm$ 130 \\
            Mixer~\cite{Hammad:2024cae} & - & 0.9859 & 416 & - \\
            MIParT~\cite{Wu:2024thh} & 0.942 &  0.9868 & 505 $\pm$ 8 & 2010 $\pm$ 97 \\
            MIParT-f.t.~\cite{Wu:2024thh} & 0.944 &  0.9878 & 640 $\pm$ 10 & 2789 $\pm$ 133 \\
            L-GATr~\cite{Brehmer:2024yqw} & 0.9423 &  0.9870 & 540 $\pm$ 20 & 2240 $\pm$ 70 \\
            L-GATr-f.t.~\cite{Brehmer:2024yqw} & \textbf{0.9446} &  0.9879 & 651 $\pm$ 11 & 2894 $\pm$ 84 \\
            \hline
            \textsc{PET}~\cite{Mikuni:2025tar,Mikuni:2024qsr} & 0.938 & 0.9848 & 340 $\pm$ 12 & 1318 $\pm$ 39\\
            \textsc{OmniLearn}~\cite{Mikuni:2025tar,Mikuni:2024qsr} & 0.942 & 0.9872 & 568 $\pm$ 9 & 2647 $\pm$ 192 \\
            \hline
            \textsc{PET v2-s} & 0.9427 & 0.987 & 505 $\pm$ 14 & 2167 $\pm$ 153\\
            \textsc{OmniLearned-s} & 0.944 & 0.9875 & 565 $\pm$ 12 & 2637 $\pm$ 128 \\
            \textsc{PET v2-m} & 0.9423 & 0.987 & 482 $\pm$ 11 & 1861 $\pm$ 61\\
            \textsc{OmniLearned-m} & \textbf{0.944 }& \textbf{0.9880} & 656 $\pm$ 12 & 3208 $\pm$ 176 \\
            \textsc{OmniLearned-l} & \textbf{0.944}& \textbf{0.9880} & \textbf{688 $\pm$ 9} & \textbf{3486 $\pm$ 157} \\
	\noalign{\smallskip}
	\end{tabular}
\end{table}

\begin{table}[th]
    \centering
    \caption{Comparison between the performance reported for different classification algorithms on the quark and gluon dataset. The uncertainty quoted corresponds to the standard deviation of nine trainings with different random weight initialization. If the uncertainty is not quoted then the variation is negligible compared to the expected value. Bold results represent the algorithm with highest performance. }
    \label{tab:results_qg}
	\begin{tabular}{lccccc}
    \noalign{\smallskip}\hline
          &  Acc &AUC & \multicolumn{2}{c}{1/$\epsilon_B$} \\
          \cline{4-5}
          & & & $\epsilon_S = 0.5$ & $\epsilon_S = 0.3$ \\
            \hline
            P-CNN~\cite{Qu:2019gqs} & 0.827 & 0.9002 & 34.7 & 91.0 \\
            PFN~\cite{Komiske:2018cqr} & - & 0.9005 & 34.7$\pm$0.4 & - \\
            ParticleNet~\cite{Qu:2019gqs} & 0.840 & 0.9116 & 39.8$\pm$0.2 & 98.6$\pm$1.3\\
            rPCN~\cite{Shimmin:2021pkm} & - & 0.9081 & 38.6 $\pm$ 0.5 & - \\
            ParT~\cite{Qu:2022mxj} & 0.840 & 0.9121 & 41.3 $\pm$ 0.3 & 101.2 $\pm$ 1.1 \\
            ParT-f.t.~\cite{Qu:2022mxj} & 0.843 &  0.9151 & 42.4 $\pm$ 0.2 & 107.9 $\pm$ 0.5 \\
            \hline
            \textsc{PET}~\cite{Mikuni:2025tar,Mikuni:2024qsr}  & 0.837 & 0.9110 & 39.92$\pm$0.1 & 104.9 $\pm$ 1.5\\
            \textsc{OmniLearn}~\cite{Mikuni:2025tar,Mikuni:2024qsr} & 0.844 & 0.9159 & \textbf{43.7$\pm$0.3} & 107.7 $\pm$ 1.5 \\
            \hline
            \textsc{PET v2-s} & 0.842 & 0.9137 & 41.7 $\pm$ 0.4 & 104.4 $\pm$ 1.3\\
            \textsc{OmniLearned-s} & 0.844 & 0.9153 & 42.9 $\pm$ 0.2 & 108.0 $\pm$ 0.5 \\
            \textsc{PET v2-m} & 0.841 & 0.9135 & 41.3 $\pm$ 0.6 & 103.8 $\pm$ 1.1\\
            \textsc{OmniLearned-m} & \textbf{0.845} & \textbf{0.9162} & 43.2 $\pm$ 0.1 & \textbf{111.2 $\pm$ 1.5} \\
	\noalign{\smallskip}
	\end{tabular}
\end{table}

The initial evaluation of \textsc{OmniLearned} on jet classification is carried out using two widely-used benchmark datasets in collider physics: top quark tagging~\cite{Kasieczka:2019dbj} and quark/gluon~\cite{Komiske:2018cqr} classification. In the top quark tagging dataset, events are simulated using \textsc{Pythia} 8 and \textsc{Delphes} with the ATLAS detector configuration. The background process consists of non-resonant jets produced via QCD and the signal consists of top quark pair production with all-hadronic decays.  Jet constituents are clustered using the anti-$k_t$ algorithm with $R=0.8$ and  all jets in the range $550$~GeV~$< p_T < 650$~GeV and $|\eta| < 2$ are saved. While multiple datasets in the pre-training data use top quarks and QCD jets, this particular dataset uses a different detector configuration and narrower $p_T$ window.  The quark/gluon dataset consists of stable particles, without detector simulation, clustered into jets. Neutrinos are excluded and the anti-$k_t$ algorithm with radius R = 0.4 is used. The quark-initiated sample (signal) is generated using a Z($\nu\nu$) + $q$ while the gluon-initiated data (background) are generated using Z($\nu\nu$) +$g$ processes. Both samples are generated using \textsc{Pythia 8}. Jets are required to have transverse momentum $\pt \in [500,550]$ GeV and rapidity $|y|<1.7$. 
\textsc{OmniLearned} is then fine-tuned on each dataset and compared  against the results obtained by the original \omni model and multiple alternatives. The results for different quality metrics are listed in Tables~\ref{tab:results_top} and \ref{tab:results_qg}. 

Notably, the improved \textsc{PET v2} model shows significantly better results compared to the previous iteration, matching and sometimes surpassing state-of-the-art models also trained from scratch on the same dataset. After fine-tuning, the performance of \textsc{OmniLearned} increases considerably and surpasses all previous benchmarks. Moreover, we observe a strong relationship between the model size of \textsc{OmniLearned} and the fine-tuning performance, with bigger models showing better results after fine-tuning. We also provide a visualization of the performance obtained by different algorithms in Fig~\ref{fig:top_tagging} for the background rejection at 30$\%$ signal efficiency.

The last classification task we investigate uses the dataset released by the ATLAS Collaboration for flavor tagging~\cite{ATLAS:2025dkv, ATLAS_JetSet_2025}. Smaller radius jets with R = 0.4 are clustered using the anti-$k_t$ algorithm and simulated using top quark decays from \textsc{POWHEG BOX}~\cite{Alioli:2010xd,Frixione:2007vw,Nason:2004rx,Frixione:2007nw} interfaced with \textsc{Pythia 8} at next-to-leading-order accuracy. Reconstructed tracks are used as inputs for the training. Jets are matched to hadron-level jets if they are found in a radius of 0.3 and are assigned the parton flavor as a label. Four different classes are considered for the classification task: b(ottom)-jets, c(harm)-jets, light jets, and jets produced through hadronic decays from tau leptons.  Besides the basic kinematic quantities used by \textsc{OmniLearned}, additional vertex information is available. To promote a fair comparison with current flavor tagging algorithms used by the ATLAS Collaboration, we include all features that are also used by the Collaboration's current state if the art GN2 architecture, with detailed explanation for each feature presented in Ref.~\cite{ATLAS:2025dkv}. In addition to the main classification task, GN2 also includes auxiliary tasks that help improve the performance of the tagger. These include the prediction of the track origin and the identification of tracks sharing a common vertex. The first auxiliary task assigns a label to each track, similar to the PID information present in other datasets. When training \textsc{OmniLearned}, we perform the jet classification using the classification  but re-purpose the generation head to perform track classification, thus enabling the model to benefit from the additional auxiliary task. In this case, all the weights from the generation head are loaded during the fine-tuning process while the output layers are replaced to match the number of classes present in each classification task. Even though the diffusion task is different than the track classification we observe the use of the pre-trained weights to still be beneficial compared to starting from random weights.

All models are trained using the full dataset consisting of 168 million jets and evaluated using the medium dataset consisting of 25.6 million jets. From the trained classifiers, the output score of the network is used to determine the tagging response. Following the ATLAS Collaboration strategy, The b- and c-tagging discriminators are defined as:

\begin{equation}
\begin{aligned}
D_b &= \log \left(
\frac{p_b}{
f_{\mathrm{c}}\,p_{\mathrm{c}} + f_{\tau}\,p_{\tau} + (1 - f_{\mathrm{c}} - f_{\tau})\,p_{\mathrm{u}}}
\right), \\[6pt]
D_c &= \log \left(
\frac{p_c}{
f_{\mathrm{b}}\,p_{\mathrm{b}} + f_{\tau}\,p_{\tau} + (1 - f_{\mathrm{b}} - f_{\tau})\,p_{\mathrm{u}}}
\right).
\end{aligned}
\label{eq:classifier}
\end{equation}

with parameters $f_{\mathrm{b}} = 0.3, f_{\mathrm{c}} = 0.2, f_{\tau} = 0.01$ optimized for GN2. We build the same discriminator functions, but instead use $f_{\mathrm{b}} = 0.2$, which improves rejection against light-jets without impacting tagging performance for other classes.

Results of the training are listed in Tab.~\ref{tab:results_atlas}, where we present results for both b- and c-tagging. Results are compared  with the ones reported by the ATLAS Collaboration trained using the same open dataset for reconstructed jets with $\pt > $ 20 GeV.~\footnote{Notice that the numbers reported are based on the training performed using the open dataset only. Public results for the GN2 model use a dataset created from a combination of top quark decays and $Z'$ events, with the latter not available at the time of the writing of this manuscript.}.

\begin{table}[th]
    \caption{Comparison between the performance reported for different classification algorithms on the ATLAS flavor tagging dataset. Values reported for GN2 are obtained from the model trained on the same dataset for a fair comparison. While the results of a single model trained are displayed, we notice that the performance listed is stable against multiple runs of the same algorithms apart from GN2 whose baseline is taken from the public results.}
    \label{tab:results_atlas}
    \begin{tabular}{lcccc}
        \hline
        \multicolumn{4}{c}{b-tagging 1/$\epsilon_B$ ($\epsilon_{\text{b}} = 70\%$)} \\
        \hline
        Algorithm & c-jets & light-jets & $\tau$-jets  \\
        \hline
        GN2 &  45.5  & 1097 & 245  \\
        \textsc{PET v2-s} & 55.7  & 1512 & 409 \\
        \textsc{PET v2-m} & 63.6  & 1772 & \textbf{494} \\
        \textsc{OmniLearned}-s & 61.4  & 1726 & 459 \\
        \textsc{OmniLearned}-m & \textbf{66.5}  & \textbf{1853} & 493 \\
        \hline
        \multicolumn{4}{c}{c-tagging 1/$\epsilon_B$ ($\epsilon_{\text{c}} = 30\%$)} \\
        \hline
        Algorithm & b-jets & light-jets & $\tau$-jets \\
        \hline
        GN2 &  21.1  & 166 & 21.4  \\
        \textsc{PET v2-s} & 22.4  & 206 & 24.1 \\
        \textsc{PET v2-m} & 24.0  & 233 & 26.7 \\
        \textsc{OmniLearned}-s & 23.6  & 221 & 26.0 \\
        \textsc{OmniLearned}-m & \textbf{24.8}  & \textbf{235} & \textbf{28.4} \\
        \hline
	\end{tabular}
\end{table}

All models show improved performance compared to the GN2 baseline and \textsc{OmniLearned} shows improved performance compared to the models trained from scratch. In particular, the impact of the fine-tuning is more pronounced for \textsc{OmniLearned-s}, improving the performance compared to the model trained from scratch by up to 15$\%$. Although bigger models also show better performance, the impact of fine-tuning on performance is smaller. However, in this case, the fine-tuning requires less than half of the network updates to achieve convergence, making the fine-tuning 50$\%$ faster than training the medium model from scratch.

\begin{figure*}[ht]
    \centering
        \includegraphics[width=.49\textwidth]{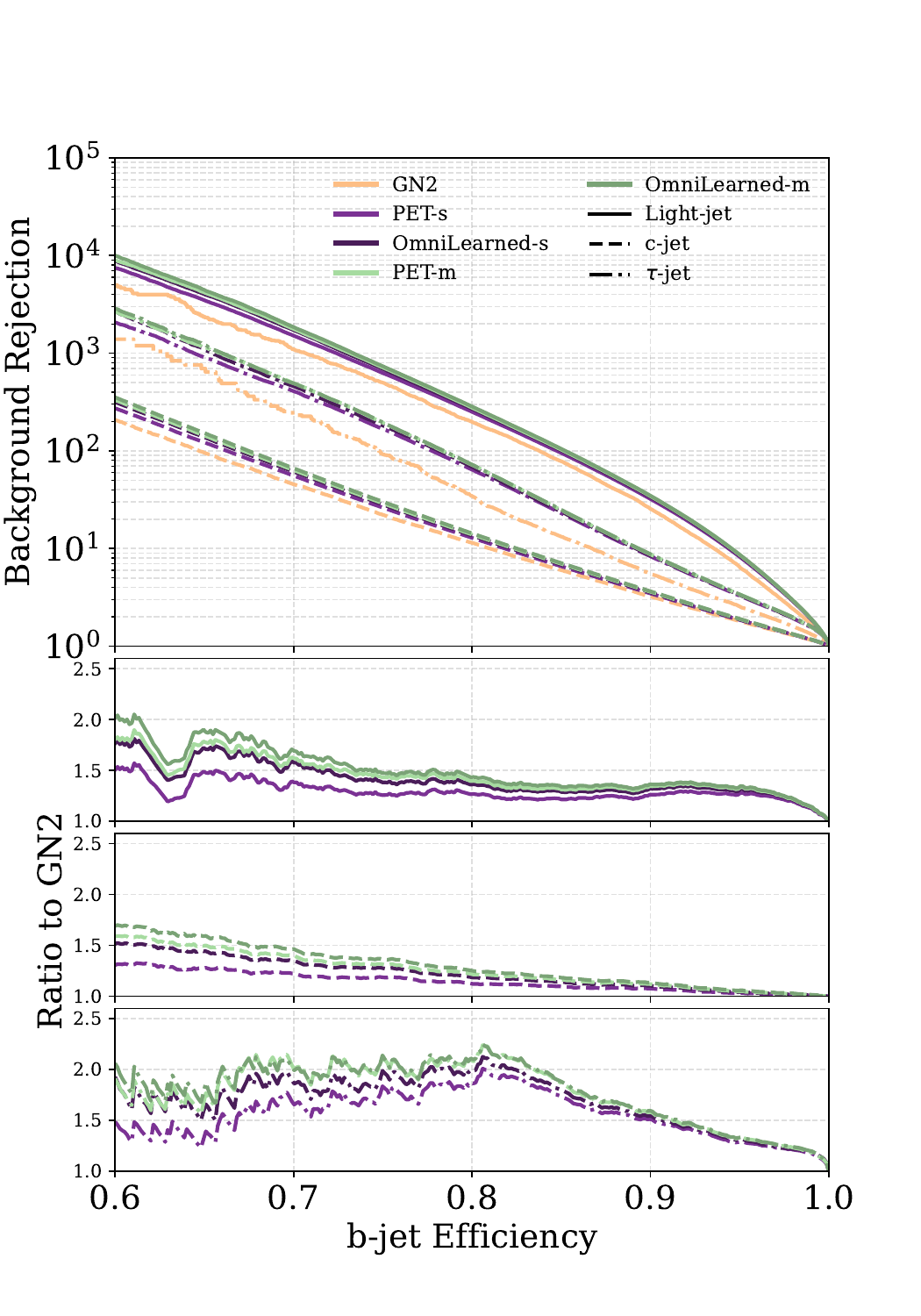}
        \includegraphics[width=.49\textwidth]{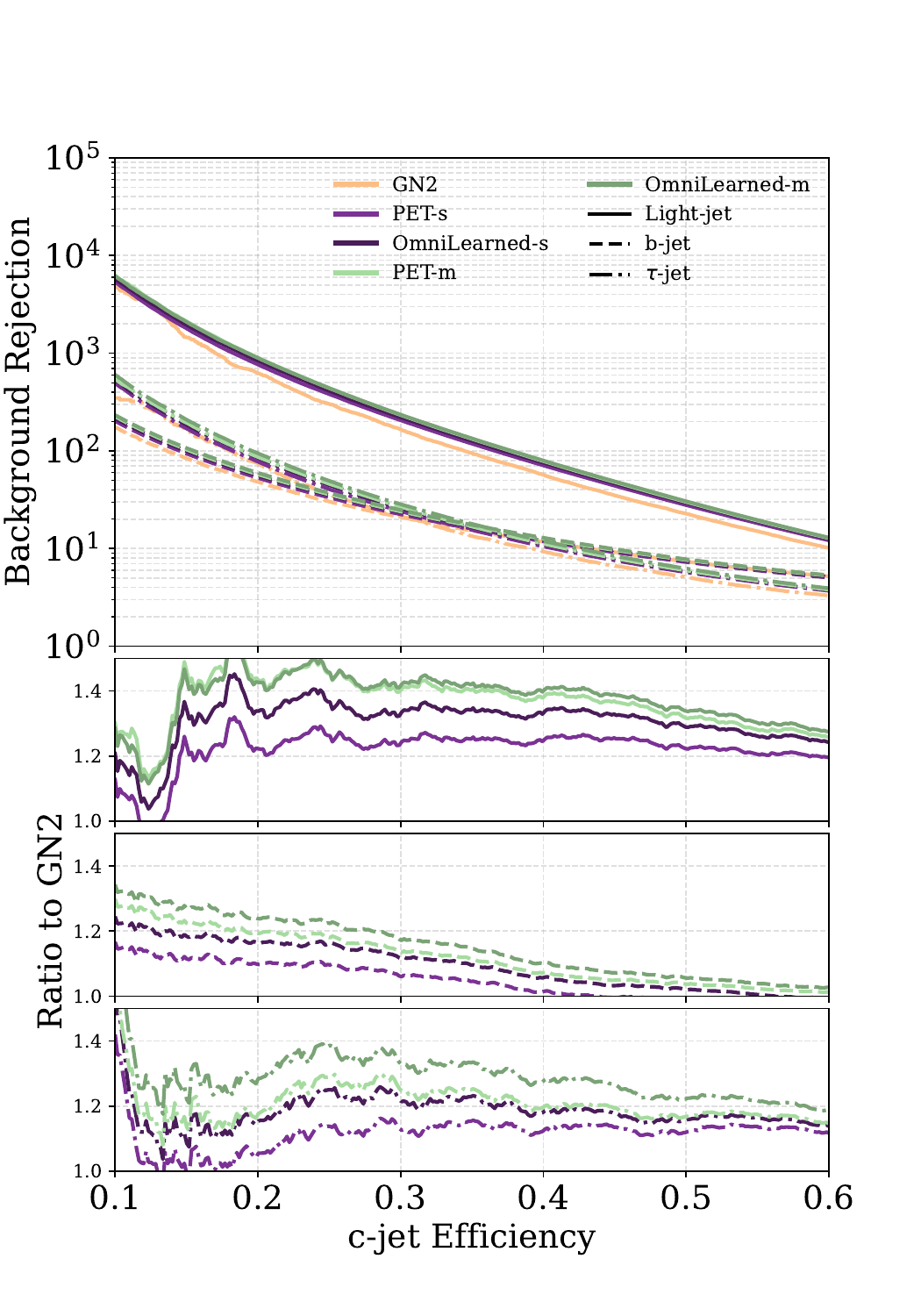}
    \caption{Receiver Operating Characteristic for b- (left) and c-tagging (right) for different algorithms. The ratio plots show the background rejection improvement compared to GN2 for different background jet classes.}
    \label{fig:roc}
\end{figure*}

We also provide the results of the Receiver Operating Characteristic (ROC) curve in Fig.~\ref{fig:roc} for different values of the signal efficiency.

We observe noticeable improvements for b- and c-tagging in terms of background rejection for all considered background types. In the case of b-tagging, light-jet rejection improves by more than 50$\%$ below 80$\%$ b-tag efficiency while $\tau$-jet rejection improves by a factor 2. Improvements to c-jet rejection are also observed to be more than 50$\%$ for b-jet efficiencies below 70$\%$. 
Improvements were also observed in c-tagging, where the light-jet rejection improves by 30$\%$ to 50$\%$ depending on the c-jet efficiency. For $\tau$- and b-jets the  rejection compared to GN2 improves by around 20$\%$ and 5$\%$-30$\%$, respectively.

\section{Anomaly Detection}
\label{sec:ad}

Our last example integrates multiple components of \textsc{OmniLearned} into one.  In particular, we consider weakly supervised, resonant anomaly detection~\cite{Metodiev:2017vrx,Collins:2018epr,Collins:2019jip,Kasieczka:2021xcg,Aarrestad:2021oeb,Karagiorgi:2021ngt}.  This is a collection of tools for automatically searching for new physics in multidimensional feature spaces.  Algorithms in this toolkit are composed of two parts: a mechanism for estimating the Standard Model background in an interval (signal region) of the reconstructed resonance mass and a procedure for determining how compatible the estimated data are with the real data.  A well-studied benchmark approach is \textsc{Cathode}~\cite{Hallin:2021wme}, which uses generative models to be able to sample background events conditioned on the reconstructed resonance mass.  This model is trained outside the signal region (in the sideband region).  Then, a classifier is trained to distinguish samples from the generative model and data in the signal region.  Historically, this classifier was trained using a small set of high-level features, but \textsc{OmniLearn} showed that sensitivity to anomalies is also possible using the full phase space if\footnote{Ref.~\cite{Cheng:2024yig,Cheng:2025ewj} showed a similar boost in sensitivity by pre-training and parameterizing the classifier in new physics properties.} the classifier uses a foundation model~\cite{Mikuni:2024qsr}.  After selecting events that the classifier flags as anomalous, a parametric fit is performed to quantify the statistical significance.  Versions of this setup have been used by ATLAS~\cite{ATLAS:2020iwa,ATLAS:2025obc} and CMS~\cite{CMS:2024nsz} to look for dijet resonances and by others to discover unisolated Upsilons with CMS data~\cite{Gambhir:2025afb} and cold stellar streams with Gaia data~\cite{Shih:2021kbt,Shih:2023jfv,Pettee:2023zra,Sengupta:2024ezl,Hallin:2025wyc}.  In this section, we deploy these methods to rediscover the top quark in the single, high-$p_T$ jet channel\footnote{Ref.~\cite{Knapp:2020dde} also used anomaly detection to rediscover the top quark, but deployed unsupervised learning and required simulations to estimate the non-top quark background.}.  Interesting in its own right as a first search with single jets and as a new benchmark to test anomaly detection methods in data with a known resonance\footnote{Ref.~\cite{Gambhir:2025afb} also provided such a dataset for the Upsilon.  Given the many challenges experienced in practice~\cite{ATLAS:2020iwa,ATLAS:2025obc,CMS:2024nsz} that were absent from phenomenology studies, it is crucial to have multiple such examples.}, this example demonstrates both the generative and discriminative capabilities of \textsc{OmniLearned}.

For this study, we use the Aspen Open Jets dataset, which contains jets collected by the CMS Collaboration during the 2016 data taking period. We consider the problem of identifying top quark initiated jets, a physics process expected to consist of only 0.1$\%$ of the jets contained in the dataset~\cite{Amram:2024fjg}. We consider jets with $\pt$ greater than 500 GeV and define the signal region consisting of jets with Soft Drop mass~\cite{Larkoski:2014wba} between 140 GeV and 220 GeV. The diffusion head in \textsc{OmniLearned} is then fine-tuned using the sidebands, including as jets with mass greater than 100 Gev and smaller than 300 GeV, while excluding the signal region. While the pre-training of \textsc{OmniLearned} does not include conditioning information, we encode the additional jet information consisting of the jet $\pt$, jet invariant and Soft Drop mass, and particle multiplicity as an additional ''particle'' to be concatenated with the real particles inside the jet after an embedding layer. The embedding layer is simply a fully-connected layer with non-linear activation function that maps the inputs and their original dimensionality to the internal dimensionality of the representation used by the model. 

\begin{figure*}[ht]
    \centering
        \includegraphics[width=.45\textwidth]{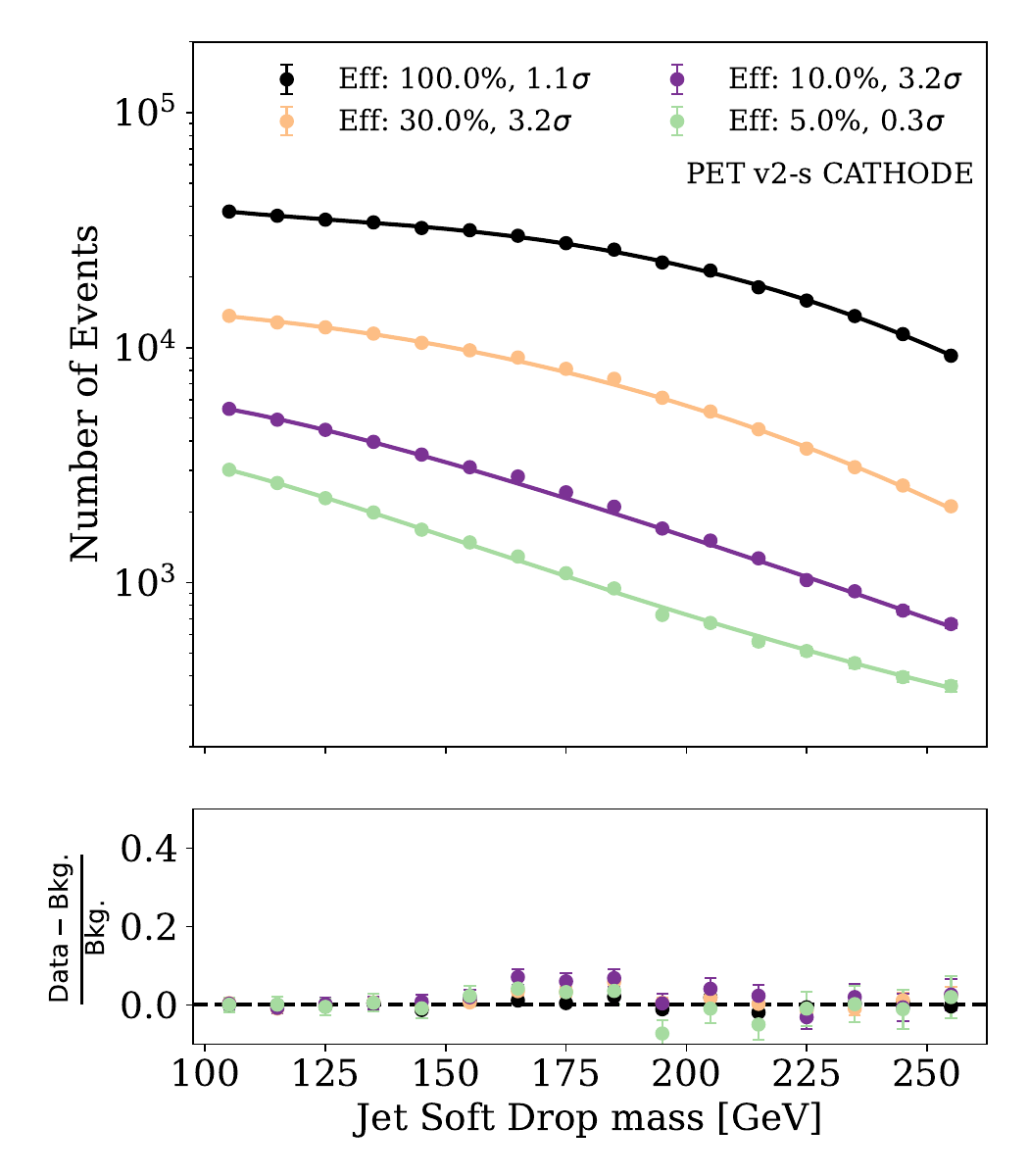}
        \includegraphics[width=.45\textwidth]{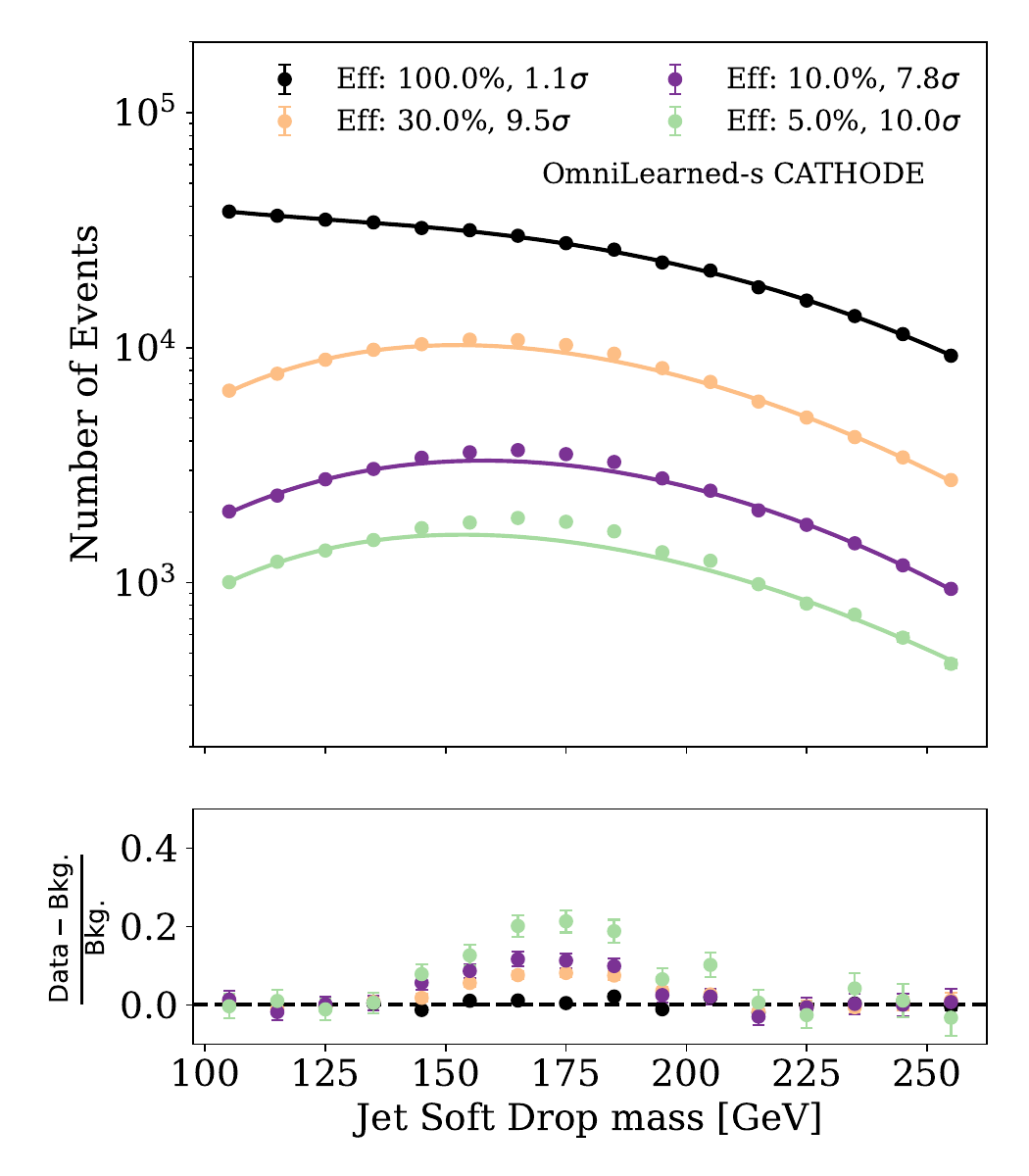}
        \includegraphics[width=.45\textwidth]{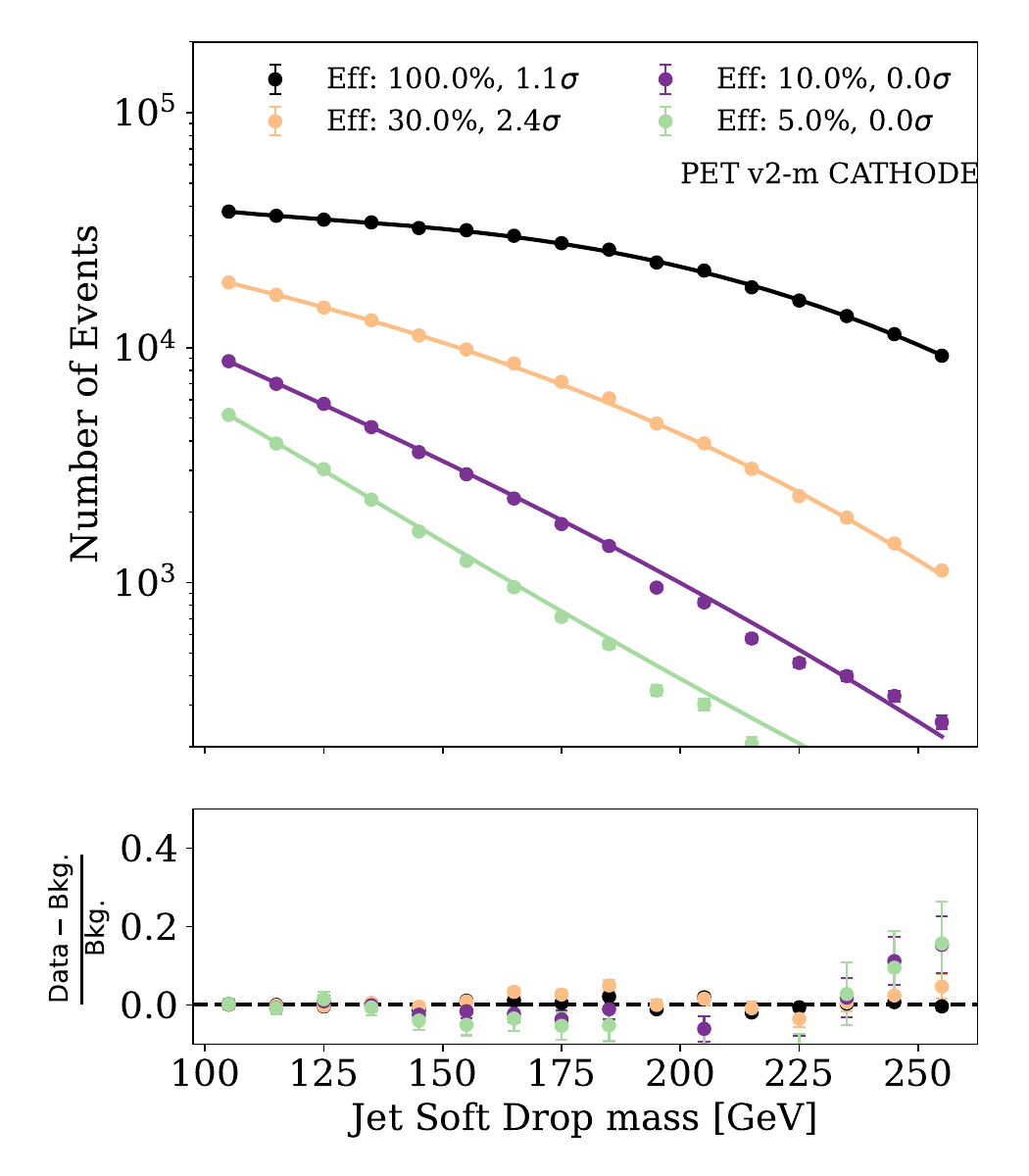}
        \includegraphics[width=.45\textwidth]{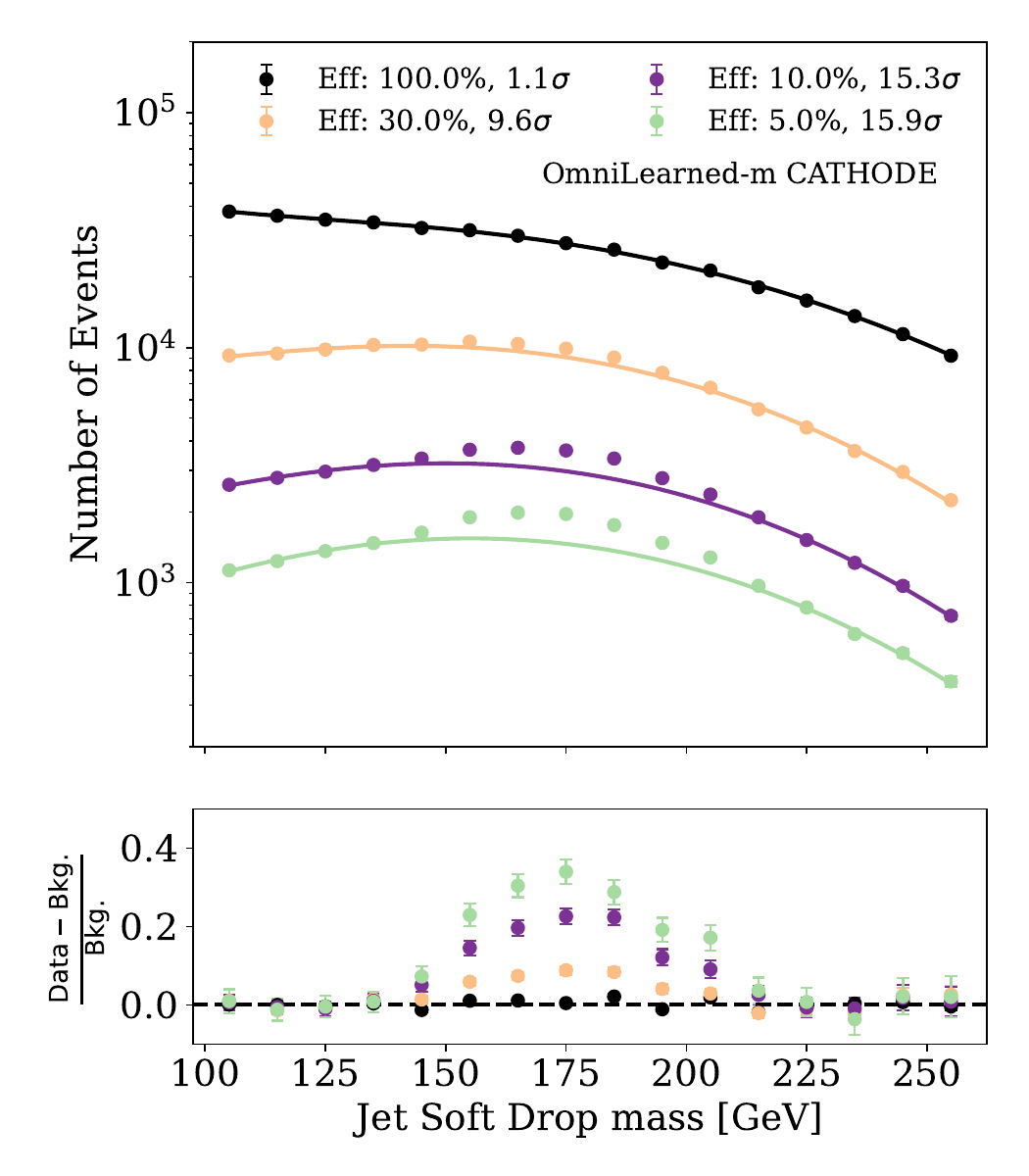}
    \caption{Anomaly detection results using the CMS Open Data. Different model sizes (rows) for models trained from scratch (left column) or fine-tuned with \textsc{OmniLearned} (right column) are shown. Different thresholds of the anomaly score, resulting in different data efficiencies, are shown together with the expected sensitivity.}
    \label{fig:ad_cathode}
\end{figure*}

To sample entire jets from the diffusion model, we first need to generate the conditioning information.  To do this, we train a second generative model using the sidebands. First, we estimate the distribution of the Soft Drop mass using the sidebands, modeled as a smoothly falling distribution with four free parameters, commonly used in dijet resonance searches~\cite{Aaltonen:2008dn,ATLAS:2015nsi,Khachatryan:2016ecr}, parameterized as:

\begin{equation}
    \frac{dN}{dm_{j}} = p_0\frac{\left ( 1-x \right )^{p_1}}{x^{p_2 + p_3\ln(x)}}, \hspace{3mm}x = m_{j}/ \mathrm{600~GeV},
    \label{eq:mass}
\end{equation}
with parameters $p_i$ fixed from the fit to the sidebands. New jet mass values are then sampled in the signal region after proper normalization of the estimated probability density function. The Soft Drop mass values are used to condition a second diffusion model whose task is to generate three numbers: the jet $\pt$, jet invariant mass, and particle multiplicity. This second diffusion model is similar to the one used in Ref.~\cite{Mikuni:2024qsr}, consisting of  fully connected layers with non-linear activations to encode the time, conditioning mass values, and the jet kinematic inputs. The full generation then proceeds as follows: first Soft Drop mass values are generated in the signal region following the distribution obtained by Eq.~\ref{eq:mass}, then the diffusion model generates the jet level distributions, which in turn are used as conditions, together with the sampled Soft Drop masses, to generate background events in the signal region with the fine-tuned \textsc{OmniLearned} model. After the background estimation, we also use \textsc{OmniLearned} to fine-tune the classifier used to distinguish data from the background prediction and create the anomaly score. Anomaly detection results are shown in Fig~\ref{fig:ad_cathode} for different model sizes, data efficiencies for the classifier selection,  and the choice of using the pre-trained model \textsc{OmniLearned} or training a model from scratch using the same datasets. The expected global significance is also listed, estimated as the ratio:

\begin{equation}
    S = \frac{N - B}{\sqrt{N}},
\end{equation}
where the number of background events $B$ is determined from the fit function (Eq.~\ref{eq:mass}) while $N$ corresponds to the total number of events in the full mass range. If the number of expected background events matches or exceeds the number of data points, the significance is set to zero.  We note that this significance is only approximate - it would be interesting to extend it\footnote{This would include a study of the false positive rate and the calibration of $p$-values.  One could also use simulations to compare with the known top quark production cross section.}, possibly with strategies to remove cuts~\cite{Gambhir:2025afb}, in the future along with scanning the mass window to promote the benchmark dataset into a full search for new physics.

The models trained from scratch show only marginal sensitivity for different anomaly score thresholds, with the medium model not able to detect the signal (consistent with Ref.~\cite{Buhmann:2023acn}). On the other hand, \textsc{OmniLearned} is  able to successfully identify the signal above discovery threshold, with the medium model yielding higher significance.

\begin{figure*}[ht]
    \centering
        \includegraphics[width=.45\textwidth]{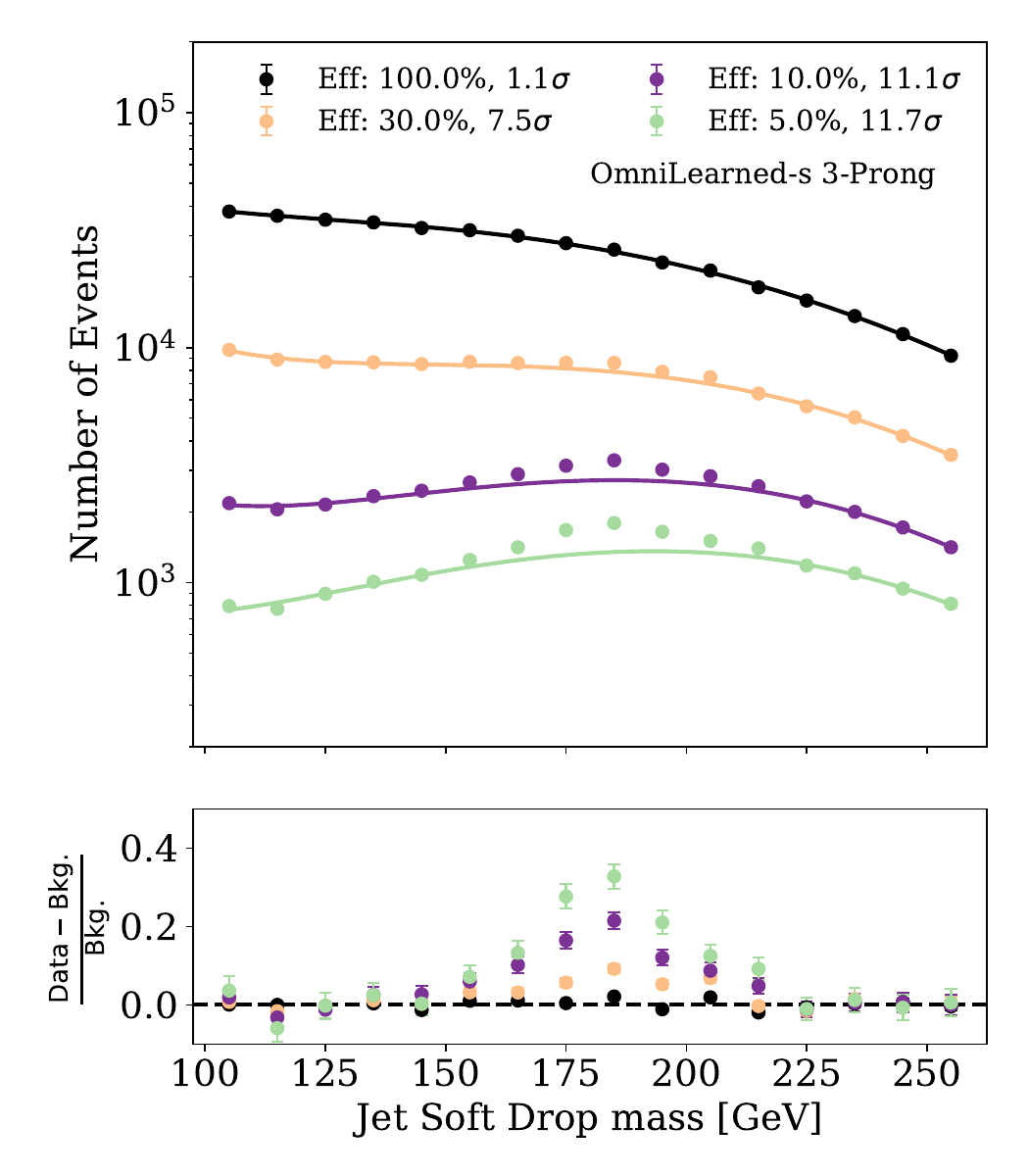}
        \includegraphics[width=.45\textwidth]{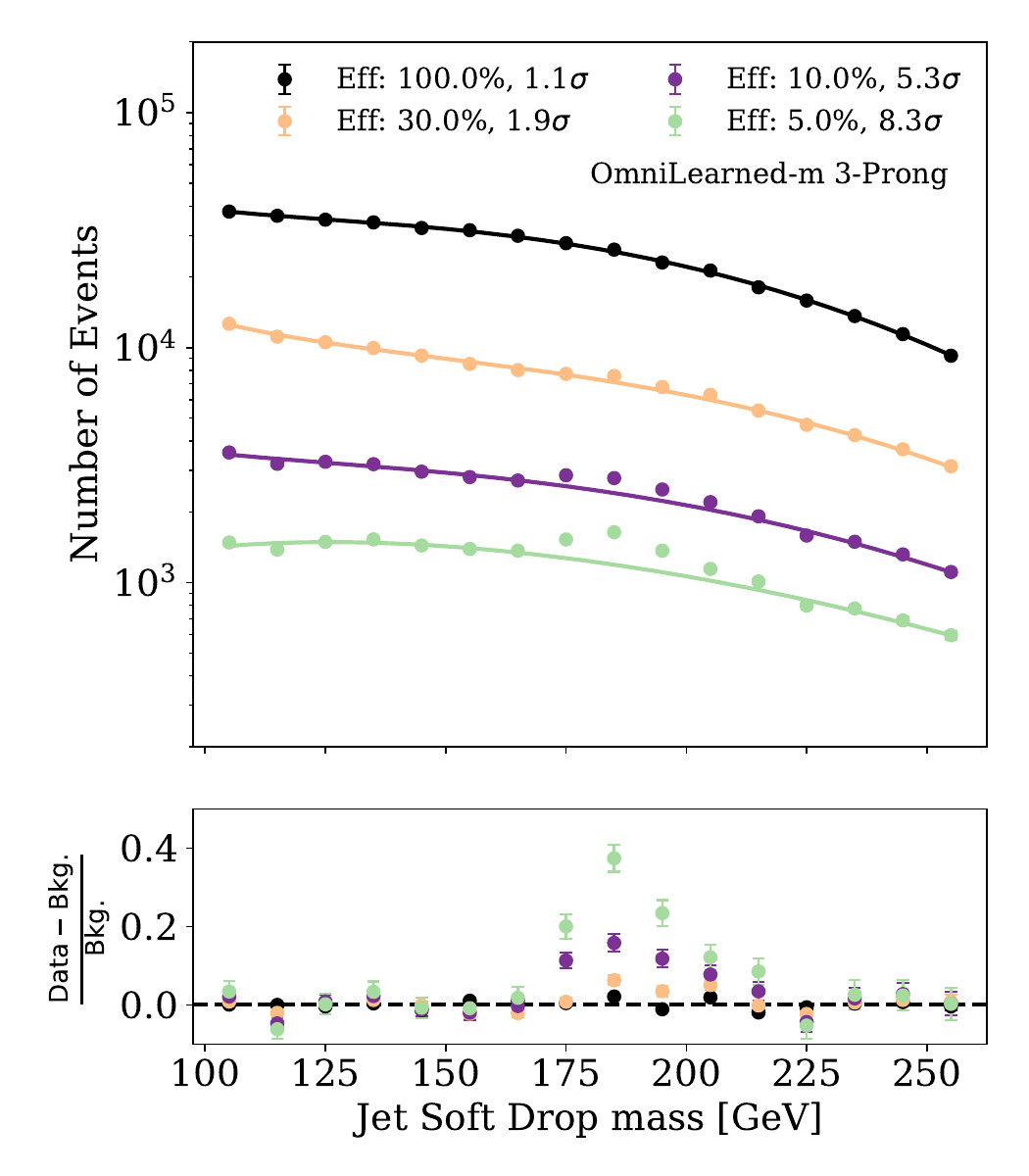}
    \caption{Anomaly detection results using the CMS Open Data. The anomaly score is calculated directly from the classes used to pre-train \textsc{OmniLearned} using the ratio between the prediction for 3-prong decays to QCD. Results for the small (left) and medium (right) models are shown.}
    \label{fig:ad_tagging}
\end{figure*}

The structure of \textsc{OmniLearned} allows for an alternative to \textsc{Cathode}-style methods.  In particular, we can leverage the multiple classes used during the training of \textsc{OmniLearned} to directly search for anomalies. For example, we can use the classes associated to generic 3-prong (the top quark decays into three quarks via an intermediate $W$ boson) decay modes from the pre-training dataset, divided by the QCD prediction nodes as the anomaly score. Notice that even though \textsc{OmniLearned} has dedicated output nodes for top quarks, we use only generic 3-prong decays without top quarks for this anomaly detection exercise.
This strategy is similar to the one proposed in~\cite{Li:2024htp}.  Even though this approach does not converge to the supervised classier with increasing data (asympotically optimal~\cite{Nachman:2020lpy}), it may still be effective for a broad set of anomalies and the theory prior may help its performance at low signal fraction~\cite{Cheng:2024yig}.
The sidebands are again used to determine the background distribution and evaluate the anomaly detection performance of the 3-prong classifier in the signal region. Results for this approach are shown in Fig.~\ref{fig:ad_tagging}.

For all model sizes the signal component is clearly visible. However, bigger model sizes lead to a small degradation of the model sensitivity. This is likely due to imperfect simulation details used during the pre-training. This application directly uses \textsc{OmniLearned} without additional fine-tuning, resulting in faster deployment.


\section{Conclusion and Outlook}
\label{sec:conclusions}

In this paper, we have introduced the follow up to the \textsc{OmniLearn} model named \textsc{OmniLearned}, a general model architecture capable of learning a useful representation of jets that is easily adaptable to many downstream tasks and datasets. With the new model, we also introduce a machine-learning ready dataset consisting of 1 billion jets that are accessible from the software package. Before pre-training, the \textsc{PET v2} model already shows improved performance compared to the previous iteration, achieving competitive results for jet tagging. The pre-training of \textsc{PET v2} using the new combined dataset leads to the \textsc{OmniLearned} foundational model, capable of achieving superior performance in traditional benchmarks for jet classification, such as the community top tagging, while also pushing the performance in flavor tagging with realistic datasets from the ATLAS Collaboration. We also show that task-specific heads of \textsc{OmniLearned} can be successfully repurposed for different tasks, such as the use of the generative head to perform track origin prediction in the ATLAS flavor tagging dataset.

The same model is used for anomaly detection, this time applied to experimental data collected by the CMS Collaboration - the first foundation model applied to real collider data to the best of our knowledge.  This application displays generalization power across detectors even when all experimental conditions are considered. In this case, we investigate two strategies to identify anomalies.  First, we perform an extended, full-phase space bump-hunt, where the background prediction is made using the generative capability of \textsc{OmniLearned} and an anomaly score is defined using the classifier capability of \textsc{OmniLearned}. We also explore an alternative option to leverage the multiple classes used to pre-train \textsc{OmniLearned} to create the anomaly score directly. This strategy has the benefit of directly using the foundational model without any additional fine-tuning. Both strategies are able to successfully identify the anomaly in the dataset, due to Lorentz-boosted top quarks, expected to be about 0.1$\%$ of the total events. 

All model implementations and pre-trained checkpoints for all model sizes are provided in the software package. Bigger models are shown to yield the best performance, both in terms of pre-training representation, evidenced by the anomaly detection results, as well as in terms of fine-tuning performance, seen by the different classification tasks. However, fine-tuning the bigger models requires significant computing resources, making the smaller models more efficient and still more performant compared to models trained from scratch. 

While we introduce \textsc{OmniLearned} as a foundational model for jet physics, we strongly believe that the methodology, and possibly even the pre-trained weights, can be useful beyond jet physics. This includes the use of full event topologies and different collision systems, leading to exciting possibilities for future investigations.

\section*{Code Availability}

The code and data for this paper can be found at~\cite{vinicius_mikuni_2026_18489564} \url{https://github.com/ViniciusMikuni/OmniLearned}.

\section*{Acknowledgments}
We thank Pradyun Hebbar, Joschka Birk, Runze Li, and Ibrahim Elsharkawy for helpful comments and testing the software environment during development. We also thank Kevin Grief for discussion on the use of the ATLAS flavor tagging dataset.
We thank our colleagues from the H1 Collaboration for allowing us to use the simulated MC event samples. We also thank DESY-IT and the MPI f\"ur Physik for providing computing infrastructure and supporting the data preservation project of the HERA experiments.
VM is supported by JST EXPERT-J, Japan Grant Number JPMJEX2509.
BN is supported by the U.S. Department of Energy (DOE), Office of Science under contract DE-AC02-76SF00515.  This research used resources of the National Energy Research Scientific Computing Center, a DOE Office of Science User Facility supported by the Office of Science of the U.S. Department of Energy under Contract No. DE-AC02-05CH11231 using NERSC awards ERCAP0034229, HEP-ERCAP0021099 and HEP-ERCAP0028249.

\appendix

\section{Model Sizes}
\label{app:models}
In the main text we introduce the \textsc{OmniLearned} models using different model sizes. The main differences between each model size is the number of transformer layers, the size of the internal representation of the model, and the number of heads considered in the transformer blocks. The values used in this work are listed in Tab.~\ref{tab:models}

\begin{table}[h!]
    \centering
    \begin{tabular}{l|ccc} \toprule
        Parameter     & Small & Medium  & Large  \\ \midrule
        Transformer Blocks    & 8  & 12 & 28\\
        Transformer Heads & 8 & 16 & 32\\
        Latent Dimension & 128 & 512 & 1024\\
        Trainable Weights & 3M & 58M & 423M\\
 
        \bottomrule
    \end{tabular} 
   \caption{Hyperparameters used in \textsc{OmniLearned}}
    \label{tab:models}
\end{table}

\bibliography{HEPML,other}
\bibliographystyle{apsrev4-1}

\clearpage

\end{document}